\documentclass{iau}

\usepackage{amsmath, bm}
\usepackage{graphicx}
\usepackage{multirow}
\usepackage{amsxtra}
\usepackage{txfonts}
\usepackage{mathpartir}
\begin{document}

\lefttitle{Shilong Liao}
\righttitle{Characterizing the astrometric quality of AGNs in Gaia-CRF3}

\jnlPage{1}{7}
\jnlDoiYr{2021}
\doival{10.1017/xxxxx}

\aopheadtitle{Proceedings IAU Symposium}
\editors{C. Sterken,  J. Hearnshaw \&  D. Valls-Gabaud, eds.}

\title{Characterizing the astrometric quality of AGNs in Gaia-CRF3}
\author{Shilong Liao$^{\ast,1,2}$ , Qiqi Wu$^{1,2}$, Ye Ding$^{1,2}$, Qi Xu$^{1,2}$,  Zhaoxiang Qi$^{1,2}$}
\affiliation{$^1$ Shanghai Astronomical Observatory, Chinese Academy of Sciences, 80 Nandan Road, Shanghai 200030, PR China \email{$^\ast$shilongliao@shao.ac.cn}\\
$^2$ University of Chinese Academy of Sciences, Beijing 100049, PR China}
\begin{abstract}
Active Galactic Nuclei (AGNs), owing to their great distances and compact sizes, serve as fundamental anchors for defining the celestial reference frame. With about 1.9 million AGNs observed in Gaia DR3 at optical precision comparable to radio wavelengths, Gaia provides a solid foundation for constructing the next-generation, kinematically non-rotating optical reference frame. Accurate assessment of systematic residuals in AGN astrometry is therefore crucial. In this talk, we analysed the parallaxes and proper motions of Gaia DR3 AGNs to characterize systematic errors and their correlations with various physical and observational properties. A subset of Gaia-CRF3 AGNs exhibits significant astrometric offsets, mainly arising from dual or lensed quasars whose structural variations induce photocenter jitter, mimicking parallax and proper motion. Such sources must be carefully excluded from reference frame construction. To this end, we introduce an astrometric quality index for each source to quantify its astrometric reliability. The results reveal a strong correlation between lower quality index values and increasing errors in position, proper motion, and parallax, demonstrating that the proposed index provides an effective metric for selecting high-fidelity AGNs as primary reference sources.
\end{abstract}


\maketitle

\section{Introduction}

Since the discovery of the first quasar in 1963, these extremely distant active galactic nuclei (AGNs) have gradually become central objects of astronomical study. In the field of astrometry, the large and well-distributed population of quasars provides an ideal foundation for constructing celestial reference frames (\cite{ma1997realization}; \cite{ klioner2022gaia}). Because of their immense distance and compact apparent structure, quasars—also known as QSOs—serve as excellent fiducial points for realizing such reference systems. With the Gaia mission observing an unprecedented number of QSOs in the optical domain and achieving an astrometric precision comparable to that obtained in the radio regime, it has become feasible to establish the next-generation, kinematically non-rotating celestial reference frame based primarily on Gaia’s measurements of these QSOs.

The Gaia Early Data Release 3 (Gaia EDR3) provides astrometric and photometric data for more than 1.8 billion celestial sources, covering a G-band magnitude range from approximately 3 to 21. These measurements are based on observations obtained by the Gaia satellite during its first 34 months of routine operations, from 2014 July 25 to 2017 May 25 (\cite{lindegren2021gaia}). The data include highly precise positions, parallaxes, and proper motions derived from a global astrometric solution. To construct a reliable sample of quasars suitable for defining the celestial reference frame, the full EDR3 catalog was systematically cross-matched with 17 external catalogs of quasars and active galactic nuclei (AGN), such as the Large Quasar Astrometric Catalog, the Sloan Digital Sky Survey quasar catalog, and the ICRF3 sources ( \cite{ klioner2022gaia}). Following this cross-identification, additional astrometric criteria were applied to ensure the extragalactic nature of the selected sources: only those with parallaxes and proper motions consistent with zero within five times their formal uncertainties were retained. As a result, a total of 1,614,173 sources were identified as quasar-like objects in Gaia EDR3. These sources are publicly available in the Gaia Archive, and represent the basis for constructing a optically defined celestial reference frame aligned with the International Celestial Reference Frame (ICRF3).

A robust assessment of the systematics in Gaia EDR3 is essential. Given their near-zero parallaxes and proper motions, QSOs serve as ideal tracers for characterizing the astrometric performance of Gaia EDR3, especially the bias in parallax and proper motion ( \cite{liao2021bprobing}; \cite{liao2021aprobing}; \cite{lindegren2021bgaia}; \cite{ding2024analysis}; \cite{ding2025analysis} \cite{cantat2021characterizing}).

Further more, among these confirmed quasars contained in Gaia-CRF3, some spectroscopically identified quasars show abnormal astrometric characteristics in the Gaia high-precision astrometric observation (\cite{wu2022squab}; \cite{ji2023strange};\cite{wu2024dulag}; \cite{makarov2022quasars}). These abnormal quasars have large proper motions or significant astrometric noises, which means that they are not suitable to be used to establish the celestial reference frame (\cite{wu2023catalog} ). Therefore, it is crucial to perform a comprehensive astrometric evaluation of the quasar candidates in Gaia-CRF3 to determine their suitability as reference frame defining sources and to characterize the astrometric systematics present in the Gaia catalog.

The paper is structured as follows. Section 2 describes the AGN sample and the astrometric properties of Gaia DR3 revealed through it. Section 3 discusses AGN candidates exhibiting abnormal astrometric behavior. Section 4 presents our preliminary results on assigning an astrometric quality index to these candidates. Finally, Section 5 provides a summary and conclusions.

\section{The Astrometric Properties of Gaia EDR3 revelled by AGNs}
\subsection{AGN-like Objects Used}
After cross-matching Gaia EDR3 with 17 external quasar catalogs and applying the astrometric filtering criteria defined in Gaia EDR3, two categories of QSO-like objects were identified: 1,215,942 sources with five-parameter astrometric solutions (hereafter FPQ) and 398,231 sources with six-parameter solutions (hereafter SPQ). In addition, a compilation of known QSOs was cross-matched with Gaia EDR3 to obtain a list of confirmed quasars and AGNs. By adopting the combined selection criteria of \cite{liao2021aprobing} and the Gaia EDR3 astrometric filters (\cite{klioner2022gaia}), the likelihood of false identifications was minimized. Under these conditions, a total of 299,004 confirmed QSOs (hereafter KQCG, \cite{liao2019compilation}) were identified in Gaia EDR3, with approximately 65$\%$ located in the northern Galactic hemisphere. Among them, 209,770 sources possess five-parameter solutions. In total, there are 1913177 QSO-like objects are used.

\subsection{The Astrometric Properties of Gaia EDR3}

\begin{figure}[h]
	\centerline{\vbox to 0pc{\hbox to 6pc{}}}
		\centering
	\includegraphics[scale=.055]{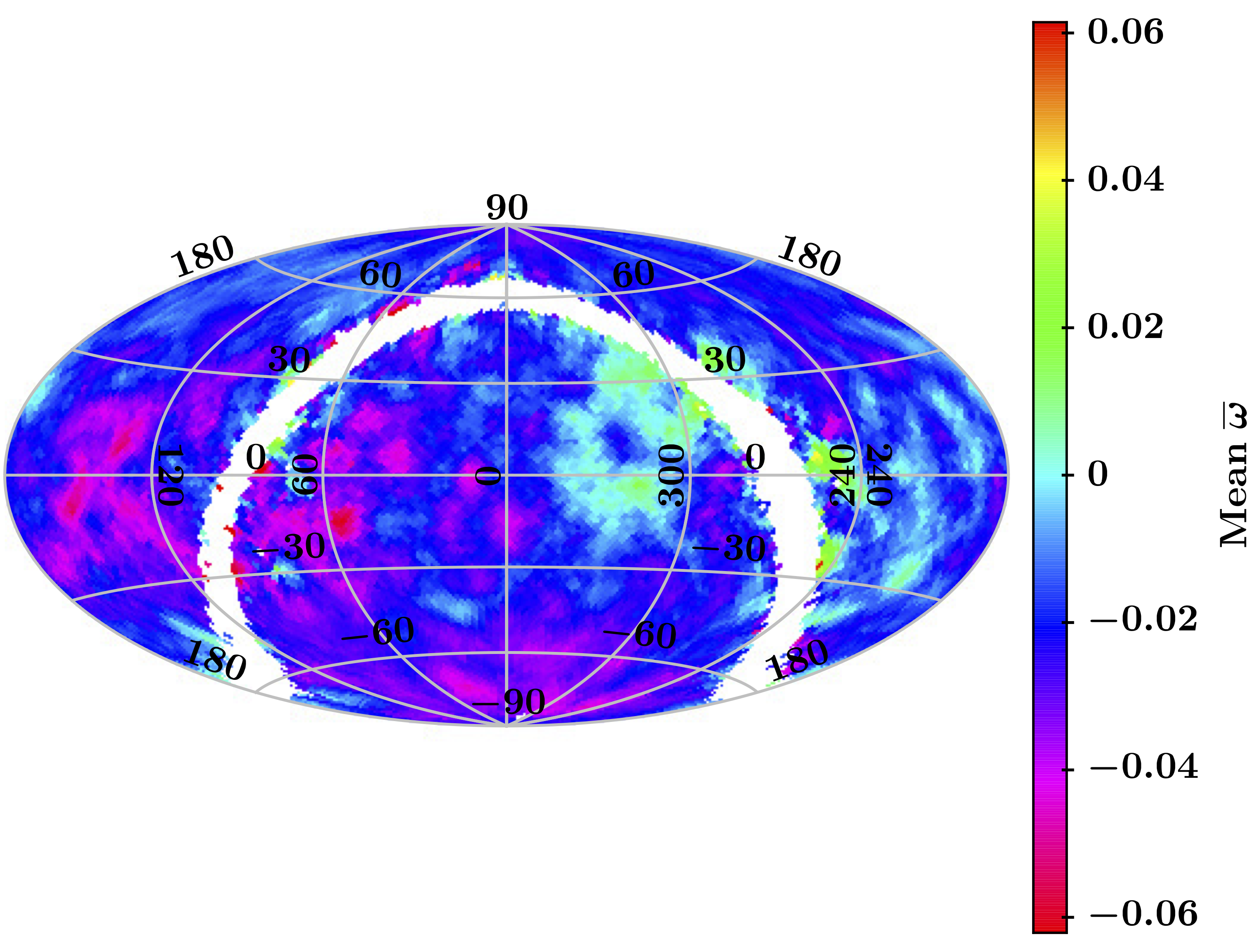}
	\includegraphics[scale=.055]{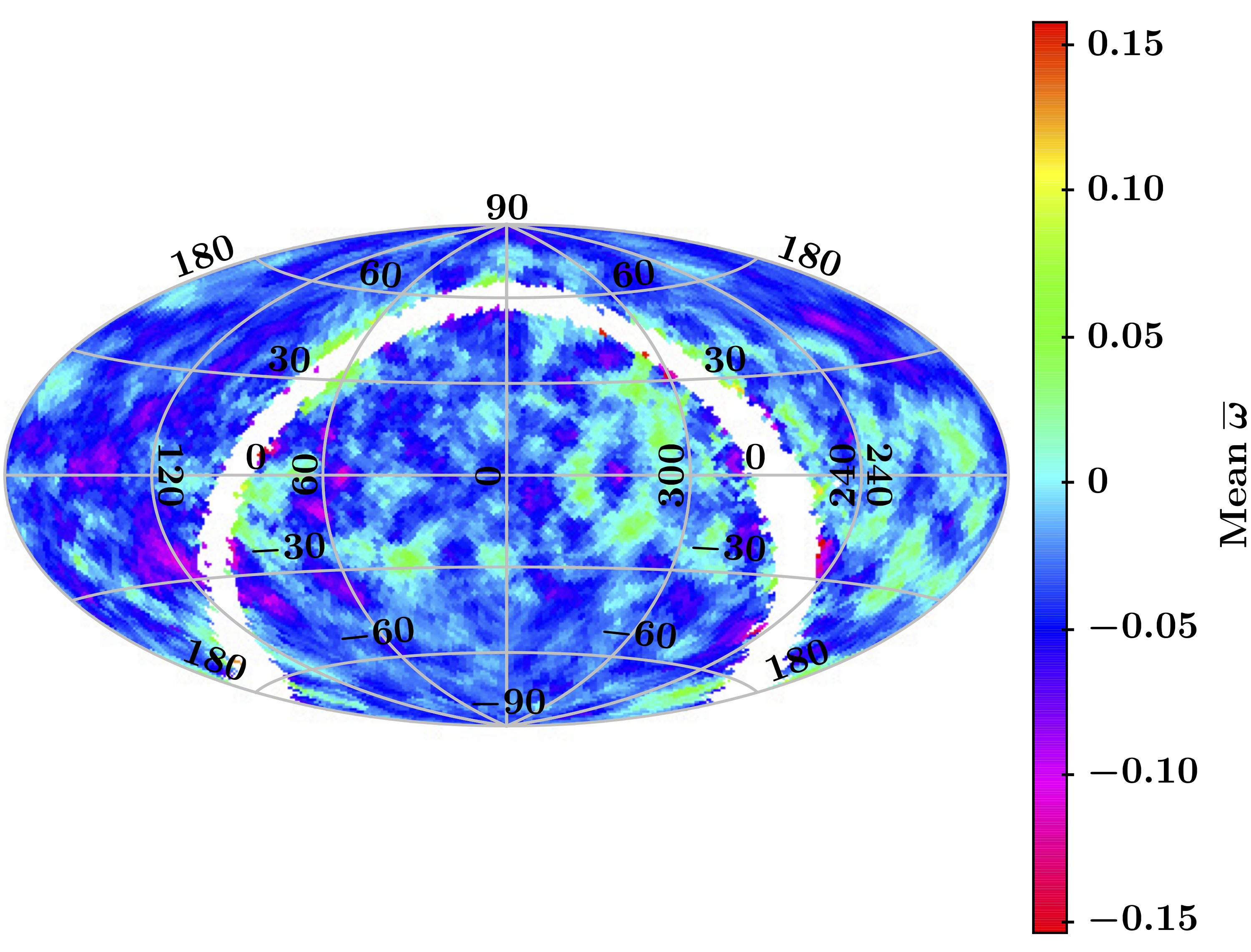}
	\caption{Maps of the generalized moving mean parallaxes (in mas) for the Gaia EDR3 quasar sample. The maps are shown in a Hammer–Aitoff projection in ecliptic coordinates. Left: Gaia EDR3 FPQ subset. Right: Gaia EDR3 SPQ subset.}
	\label{GMM_mean_plx}
\end{figure}

Since AGNs are assumed to have zero parallaxes, any nonzero parallax measured in the Gaia catalog directly reflects the parallax bias. To quantify this global bias, we computed the generalized moving mean  of the parallax values using a weighting function that accounts for the geometric correlation of neighboring sources on the celestial sphere. A negative global bias is observed in both the FPQ and SPQ samples, approximately -21.3 $\mu$as ($\sigma$ = 9.77 $\mu$as) for FPQ and -27.5 $\mu$as ($\sigma$ = 22.23 $\mu$as) for SPQ, consistent with the results of \cite{lindegren2021bgaia}. In addition, large-scale spatial variations are evident; for example, in the FPQ sample, the parallax bias reaches -40 $\mu$as within a 5° radius around ($\lambda$, $\beta$) = (60°, 10°), while it is 20 $\mu$as near ($\lambda$, $\beta$) = (300°, 20°), corresponding to a peak-to-peak variation of about 60 $\mu$as across the sky, see Figure \ref{GMM_mean_plx}. The parallaxes of EDR3 quasars show an rms amplitude of 9.9 $\mu$as spatial correlation with the angular of larger than 18$^\circ$ (\cite{liao2021aprobing}).

Similarly, for the FPQ subset, the global mean proper motions are -0.3 $\mu$as yr$^{-1}$ ($\sigma$ = 11.4 $\mu$as yr$^{-1}$) in the $\mu_{\alpha \ast}$ direction and -1.2 $\mu$as yr$^{-1}$ ($\sigma$ = 10.2 $\mu$as yr$^{-1}$) in the $\mu$$_{\delta}$ direction. For the SPQ subset, the corresponding values are -0.6 $\mu$as yr$^{-1}$ ($\sigma$ = 23.4 $\mu$as yr$^{-1}$) in $\mu_{\alpha \ast}$  and -4.2 $\mu$as yr$^{-1}$ ($\sigma$ = 25.0 $\mu$as yr$^{-1}$) in $\mu$$_{\delta}$. 

Using this large AGN sample, we further analyzed the properties of the celestial reference frame Gaia-CRF3. For quasars with five-parameter solutions, the derived rotation and glide vectors are (3, 2, 1) $\pm$ 0.5 $\mu$as yr$^{-1}$ and (0, -4, -2) $\pm$ 0.5 $\mu$as yr$^{-1}$, respectively. A slight systematic dependence of the rotation, particularly the x-component, on Gaia color and G magnitude is observed. The vector spherical harmonics  fitting results indicate that a non-rotating frame can be achieved with an appropriately selected quasar subset in Gaia EDR3. Additionally, a small rotation difference between the northern and southern hemispheres is detected, warranting further investigation in future data releases.

Systematic errors are unavoidable in Gaia’s published astrometric data. Achieving high-precision and high-accuracy parallaxes is crucial for a wide range of astronomical studies, making the investigation and correction of parallax bias essential. \cite{lindegren2021bgaia} ( hereafter L21) proposed a comprehensive correction model for Gaia EDR3 parallaxes, derived from quasars in the Gaia-CRF3, stars in the Large Magellanic Cloud, and physical binaries. The correction depends on the G-band magnitude, color, and ecliptic latitude of the sources. However, we demonstrated that the L21 correction performs poorly in the Galactic plane and proposed an alternative correction tailored for this region, achieving improved accuracy (\cite{ding2024analysis}).

\section{AGN Candidates With Abnormal Astrometric Characteristics }
Some of spectroscopically confirmed sources exhibit significantly abnormal astrometric parameters. For instance, certain quasars show large deviations in proper motion and parallax, or display noticeable positional differences between Gaia DR2 and EDR3. Such anomalies suggest that Gaia’s high-precision measurements may be sensitive to intrinsic source variability or structural jitter.  Nevertheless, these quasars are unsuitable for defining the celestial reference frame and should therefore be excluded from the reference frame candidate list.

\begin{figure}[h]
	\centering
	\includegraphics[scale=0.36]{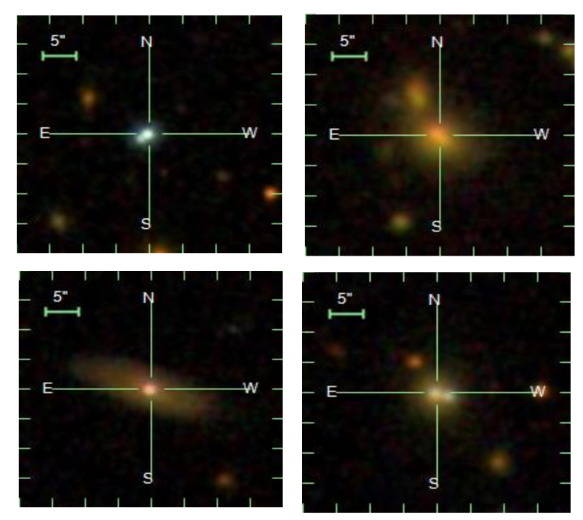}
	\includegraphics[scale=0.25]{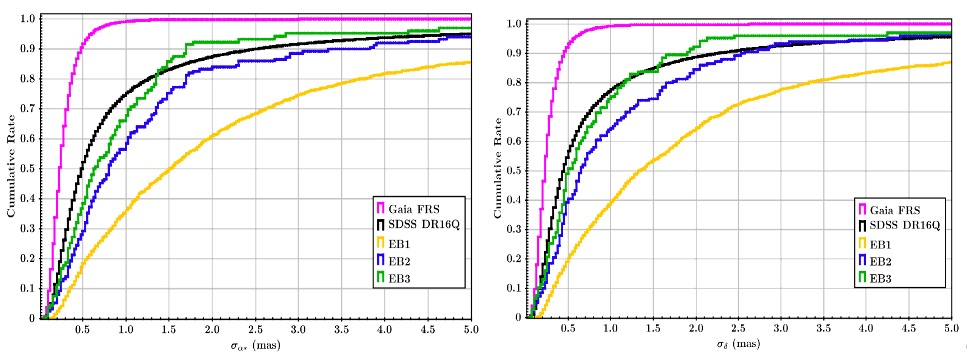}
	\caption{Four SDSS DR16 quasars images with abnormal astrometric behavior (Left panel).Cumulative distribution histograms of positional errors in right ascension (RA) and declination (Dec) for different quasar subsets. EB1–EB3 represent the subsets of abnormal (or “strange”) AGN candidates identified in this work (middle and right panels).  }
	\label{strang_agn}
	\vspace{-0.1cm} 
	\setlength{\abovecaptionskip}{1.cm}
\end{figure}

We made a preliminary attempt to identify quasars exhibiting abnormal astrometric behavior. Specifically, we selected SDSS quasars with more than one Gaia EDR3 counterpart within 1 arcsecond, as these sources are likely affected by nearby objects, resulting in increased positional uncertainties. For quasars with a single Gaia match within 1 arcsecond of the SDSS position, we further identified abnormal astrometric candidates using several Gaia astrometric parameters. Large values of these parameters indicate poor astrometric fits or unresolved double sources, and thus can be used to assess the quality of point-spread-function (PSF) fitting and the reliability of the measurements. Examination of SDSS images for these spectroscopically confirmed quasars revealed that many exhibit clearly extended morphologies, see Figure \ref{strang_agn}. Based on these findings, we proposed a set of criteria for selecting abnormal quasars using Gaia astrometric data, resulting in two catalogs containing 155 and 44 sources, respectively—both representing potential quasar pair candidates (\cite{wu2022squab}; \cite{wu2023catalog}).

Building on this work, several studies have shown that the Gaia Multipeak (GMP) method is an efficient approach for identifying active galactic nucleus (AGN) pair candidates. The number of identified pairs depends on the size of the input AGN catalog and is typically limited to spectroscopically confirmed sources. In this study, a larger and highly reliable catalog of GMP-selected AGN pair candidates was compiled from six million Gaia AGN catalog entries, most of which lack spectroscopic data. By comparing the properties of GMP AGN pair candidates with those of normal AGNs, we identified significant differences in their astrometric and multiband color distributions, which were used to establish optimized selection criteria. This led to the creation of the DUal and Lensed AGN candidate catalog (DULAG, \cite{wu2024dulag}), containing 5286 sources and a high-confidence “Golden Sample” of 1867 sources. Among these, 37 have been confirmed as dual or lensed AGNs. Reference redshifts are provided for most Golden Sample sources, and three close AGN pair candidates were identified.

We further present optical confirmations for quasars with multiple detections from the above candidate lists (\cite{ji2023strange}). Based on stellar indicators such as measurable proper motion or parallax, stellar-like spectra and color–color diagnostics, we identified 65 quasar–star pairs, as well as two dual-quasar (DQ) candidates. Additionally, 56 lensed-quasar (LQ) candidates were found with similar color characteristics, including 13 previously known LQs and 5 reported LQ candidates. A search of the HST archive yielded high-resolution imaging for 19 of these targets, among which 10 are confirmed LQs and one is a confirmed DQ.

\section{The astrometric quality index of AGNs in Gaia}
As mentioned above, with the improvement of observational precision, some quasars in Gaia-CRF3 exhibit abnormal astrometric parameters and are therefore unsuitable as reference frame sources. To identify high-precision and highly stable reference frame sources, we perform a multidimensional quantitative analysis based on color, spectrum, variability, positional accuracy, and structural morphology. A comprehensive evaluation of quasar astrometric parameters was conducted to identify and remove sources exhibiting positional instability or structural irregularities. For the first time, an astrometric quality index was established for each target to quantitatively characterize its astrometric fidelity.

\begin{table}[h!]
	\centering
	\caption{The criteria of different parameters}\label{table_astrometric}
	{\tablefont\begin{tabular}{@{\extracolsep{\fill}}lcccc}
			\midrule
			Parameter&Best source & Worst source&
			Optimal value&
			Worst value\\
			\midrule
			DESI Type&  PSF &   Other Types &  PSF &    Other Types\\
			ps1$\_$morph&     $<$0.12 &   $>$0.2&  0 &   8.1054\\
				brefactor&     $<$1 &   $>$2&  0 &   70.56\\
				Gaia solution&    5-parameter &   2/6-parameter&  5-parameter  &  2/6-parameter\\
			astrometric$\_$gof$\_$al&     $<$2&   $>$3&  0 &   1195.31\\
						astrometric$\_$excess$\_$noise&     $<$1&   $>$2&  0 &   102.4\\
						ipd$\_$gof$\_$harmonic$\_$amplitude&     $<$0.2&   $>$0.226&  0 &   6.65\\
							ipd$\_$frac$\_$multi$\_$peak&     0&   $>$7&  0 &   99\\
								ruwe&     $<$1.1 &   $>$1.2&  1 &   17.14\\
			\midrule
	\end{tabular}}
	\tabnote{\textit{Notes}: [1] DESI Type: PSF means the source is treated as point source in DESI image. [2] ps1$\_$morph: Morphological index calculated from Pan-STARRS1 observation.[3] brefactor is the value modified from bp$\_$rp$\_$excess$\_$factor provided in the Gaia catalog. [4]For the remaining parameters listed in the table, they correspond to those provided directly in the Gaia catalog. }
\end{table}

\begin{figure}[h]
	\centerline{\vbox to 0pc{\hbox to 6pc{}}}
	\centering
	\includegraphics[scale=.50]{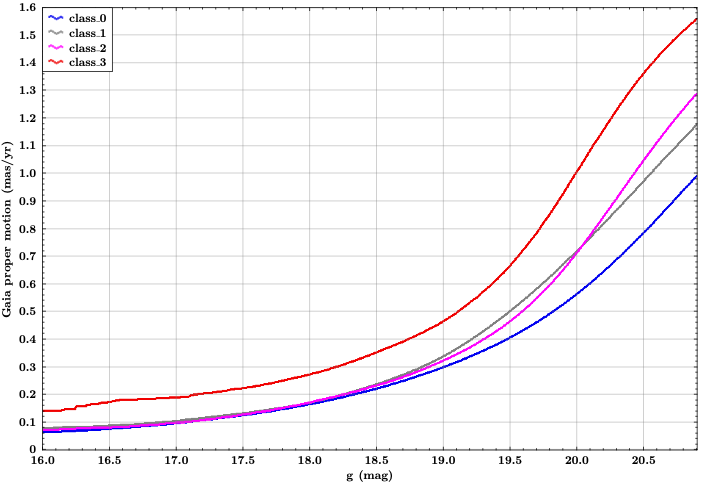}
	\includegraphics[scale=.50]{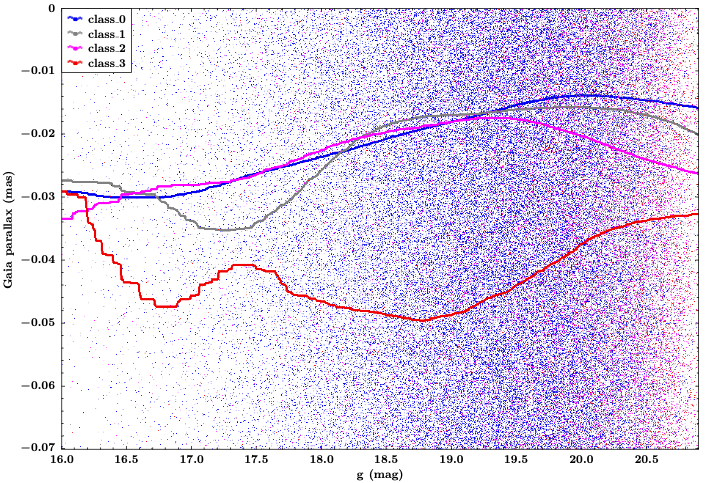}
	\caption{Proper motion (left) and parallax (right) as a function of magnitude for Gaia-CRF3 sources of different quality classes.}
	\label{Astrometric_index}
\end{figure}

The criteria for assigning an astrometric quality index of “GOOD” to a candidate are as follows: the source should exhibit a point-like morphology, possess a Gaia five-parameter astrometric solution, display low Gaia excess noise, show no internal resolvable structure, and maintain consistent morphology across different Gaia scanning directions. Conversely, AGN candidates exhibiting irregular or extended morphology, having only a Gaia two- or six-parameter solution, showing large Gaia excess noise, or presenting clear internal structure are classified as “BAD” candidates. The details of the criteria of different parameters can be found in Table \ref{table_astrometric}\footnote{More details can be found in the PhD Thesis of Wu Qiqi, titled "Research on the selection method of celestial reference frame sources for CSST"}.

Based on the criteria listed in the Table \ref{table_astrometric}, we apply the TOPSIS (Technique for Order Preference by Similarity to Ideal Solution) method for multi-parameter comprehensive evaluation. TOPSIS is a classical multi-criteria decision-making approach suitable for mixed data types. Its core concept is that the optimal solution should be closest to the “positive ideal solution” and farthest from the “negative ideal solution.” In practice, we determine the optimal and worst values for each parameter, calculate the Euclidean distance of each source to both ideal solutions, and rank them according to relative closeness. By validating the results with known good and poor samples and iteratively adjusting parameter weights, we finally obtain a comprehensive score for each source — the astrometric quality index.

Based on the distribution of the astrometric quality index, we classify the AGN candidates in Gaia-CRF3 into four classes. Their parallax and proper motion distributions as a function of magnitude are shown in the Figure \ref{Astrometric_index}. The lowest-quality group contains about 140,000 sources, accounting for approximately 9$\%$ of the total. These objects either exhibit significant extended structures, have observable internal features, or are affected by nearby bright stars. Compared with other classes, they show notably larger systematic biases in parallax and proper motion. Therefore, these sources should be used with caution to serve as high-quality reference objects for constructing the celestial reference frame.

\section{Summary and Conclusions}
In summary, our analysis of Gaia EDR3 AGNs reveals a global parallax bias of about -20 to -30 $\mu$as and small systematic trends in color and magnitude, with spatial correlations up to 10 $\mu$as over 18$^\circ$. While most AGNs provide stable astrometric solutions, a fraction show abnormal behaviors - often associated with dual or lensed systems - making them unsuitable for reference frame construction. To ensure the robustness of the optical celestial reference frame, we introduce an astrometric quality index that effectively identifies high-quality, stable AGNs as primary reference sources.


\begin{acknowledgements}
	This work has been supported by  the
	Strategic Priority Research Program of the Chinese Academy of Sciences, Grant No.XDA0350205, the National Natural Science Foundation of China (NSFC) through grants 12173069, the Youth Innovation Promotion Association CAS with Certificate Number 2022259, the Talent Plan of Shanghai Branch, Chinese Academy of Sciences with No. CASSHB-QNPD-2023-016. 
\end{acknowledgements}


\bibliography{mysample}
\bibliographystyle{iaulike}

\end{document}